\documentclass[12pt,a4paper]{article}
\usepackage[cp866]{inputenc}
\usepackage{amsmath}
\usepackage[dvips]{epsfig}
\usepackage{graphicx}
\usepackage{epsfig}
\newcommand{\be}{\begin{equation}}
\newcommand{\ee}{\end{equation}}
\newcommand{\ben}{\begin{equation*}}
\newcommand{\een}{\end{equation*}}
\newcommand{\bea}{\begin{eqnarray}}
\newcommand{\eea}{\end{eqnarray}}
\newcommand{\ar}{\begin{array}}
\newcommand{\arn}{\end{array}}

\def\pnot{\mbox{${\not{\hbox{\kern-3.0pt$p$}}}$}}
\def\qnot{\mbox{${\not{\hbox{\kern-2.0pt$q$}}}$}}
\def\enot{\mbox{${\not{\hbox{\kern-2.0pt$e$}}}$}}
\def\knot{\mbox{${\not{\hbox{\kern-2.0pt$k$}}}$}}

\def\fun#1#2{\lower3.6pt\vbox{\baselineskip0pt\lineskip.9pt\ialign
{$\mathsurround=0pt#1\hfil##\hfil$\crcr#2\crcr\sim\crcr}}}

\begin{document}
\numberwithin{equation}{section}     
\sloppy
\renewcommand{\baselinestretch}{1.0} 

\begin{titlepage}

\hskip 11cm \vbox{ \hbox{Budker INP 2018-7}  }
\vskip 3cm
\begin{center}
{\bf On the cancellation of radiative corrections
to the cross section of electron-proton scattering}
\end{center}

\centerline{V.S.~Fadin$^{a, b\,\dag}$, R.E.~Gerasimov$^{a, b\,\ddag}$}

\vskip .6cm

\centerline{\sl $^{a}$
Budker Institute of Nuclear Physics of SB  RAS, 630090 Novosibirsk
Russia}
\centerline{\sl $^{b}$ Novosibirsk State University, 630090 Novosibirsk, Russia}

\vskip 2cm

\begin{abstract}
The largest radiative corrections to the cross section of electron-proton scattering at high energies are associated with  emission of photons, real and virtual,  by electron.  They contain large logarithms coming from   soft and collinear photons. Cancellation of the contributions of the soft photons to virtual and real corrections is well known. Less known is the fact that the contributions of  the photons  collinear to scattered  electrons are also cancelled for the  most of experiments.  On the contrary, contributions of the  photons  collinear to initial  electrons as a rule  are not  cancelled.  It is shown, however, that these contributions are cancelled for the experimental set up suggested
by A.A. Vorobev for measurement of proton radius.
\end{abstract}


\vfill \hrule \vskip.3cm \noindent $^{\ast}${\it Work supported
in part by the Ministry of  Science and Higher Education of Russian Federation and
in part by  RFBR,   grant 16-02-00888.} \vfill $
\begin{array}{ll} ^{\dag}\mbox{{\it e-mail address:}} &
\mbox{fadin@inp.nsk.su}\\
^{\ddag}\mbox{{\it e-mail address:}} &
\mbox{r.e.gerasimov@inp.nsk.su}\\
\end{array}
$
\end{titlepage}

\vfill \eject

\section{Introduction}
The striking  difference in the proton radius values extracted at the Paul Scherrer Institute from  the 2S-2P transition in muonic hydrogen \cite{Pohl:2010zza, Antognini:1900ns} and obtained from electron-proton  scattering and hydrogen spectroscopy \cite{Mohr:2008fa}  (for a review, see Ref.~\cite{Bernauer:2014nqa}) caused a surge  of interest of theorists and experimentalists to the problem  and got the name "proton radius puzzle"  \cite{Pohl:2013yb, Carlson:2015jba}.  Latest electron scattering  experiments at Jlab~\cite{Zhan:2011ji} and MAMI~\cite{Bernauer:2010wm} and  hydrogen spectroscopy experiments \cite{Beyer:2017, Fleurbaey:2018fih} not only did not resolve the puzzle, but made it even more confusing.

Currently new scattering experiments are being prepared. A distinctive feature of one of them \cite{Vorobev}, which was suggested by A.A. Vorobev and  has  to  be performed with a low-intensity electron beam at MAMI,  is that instead of detecting a scattered electron, as in previous experiments, it is supposed to detect with a high precision a recoil proton in the region of low momentum transfer  squared $Q^2$ from 0.001 to 0.04 GeV$^2$. The aim is to extract the proton radius with 0.6 percent precision, which could be decisive in solving the proton radius puzzle. To this end, it is planned to achieve 0.2 percent accuracy of the cross section $d\sigma/(d Q^2)$ measurement.

Such accuracy requires precise  account of radiative corrections. Although calculation of the radiative corrections to the electron-proton scattering cross section has a long history (see, for example, Refs.~\cite{Tsai:1961zz}-\cite{Maximon:2000hm} and recent reviews  \cite{Carlson:2007sp}-\cite{Afanasev:2017gsk}), the results obtained before cannot be completely applied to the experiment discussed above. The reason it that they were obtained for experiments in which scattered electrons were detected (honestly speaking, there was the experiment \cite{Qattan:2004ht} where the recoil proton was detected; but calculation  of the radiative corrections to this experiment was not explained). Since the radiative corrections include contributions of inelastic processes with photon emission, they depend  strongly  on experimental conditions, so that the corrections calculated  for experiments with detection of scattered electrons are not suitable for experiments with detection  of recoil proton. It occurs  that the radiative corrections for experiments with detection  of recoil proton have a new unexpected and pleasant property  -- cancellation of the most important pieces. This paper is dedicated to explanation and derivation of this property.

\section{One-loop corrections}
We will denote four-momenta of initial and final electron (proton)  as $l\;$($p$) and $l'\;$($ p'$); $l^2=l'^2=m^2, \;\; p^2=p'^2=M^2$, and will use the designations $Q^2 = - q^2, \;\; q=p-p'$  both for elastic the and inelastic processes. In the following we assume  electrons to be ultrarelativistic in both the initial and final states and also neglect $m^2$ compared with $M^2$ and $Q^2$.

In the Born approximation the electron-proton cross section is known since 1950 \cite {Rosenbluth:1950yq}. It can be written as
\be
\frac{d\sigma_B}{dQ^2} = \frac{2\pi\alpha^2}{Q^4}\frac{2M(E_l^2+E'^2_l+M(E_l-E'_l)) }{E_l^2(2M+E_l-E'_l)}
{\left(\tau \,
 G_M ^ 2 (Q^ 2) + \epsilon \, G_E ^ 2 (Q^ 2) \right)}~,
 \label{born}
\ee
where $E_l$ and $E'_l$  are initial and final electron energies in the rest frame of the initial proton,
\be
\epsilon = \frac{2E_lE'_l-M(E_l-E'_l) }{E_l^2+E'^2_l+M(E_l-E'_l)}~, \label{epsilon}
\ee
$\tau =Q^2/(4M^2)$, $G_E(Q^2)$ and $G_M(Q^2)$ are electric and magnetic form factors, which are expressed through the Dirac and Pauli form factors $f_1(Q^2)$ and $f_2(Q^2)$,  defined by the  matrix element of the electromagnetic current $J^\mu$ between initial and final proton states
\be
<P'|J^\mu(0)|P> = \bar{u}(p')\,\left(f_1(Q^2)\,\gamma^\mu + f_2(Q^2)\,\frac{[\gamma^\mu, \gamma^\nu]q_\nu}{4M}\right)u(p)~,
\ee
as
\be
G_M(Q^2)=f_1(Q^2)+f_2(Q^2),\;\; G_E(Q^2)=f_1(Q^2)-\frac{Q^2}{4M^2}f_2(Q^2)~.
\ee

With elastic scattering $E'_l$ and $Q^2$ are not independent variables and are expressed through the initial electron energy $E_l$ and the scattering angle $\theta$ in the proton rest frame:
\be
E'_l=\frac{E_l}{1+(2E_l/M) \sin ^2 (\theta/2)}, \;\; Q^2 = 4E_lE'_l \sin ^2
(\theta/2)~.  \label{E'l Q2}
\ee
The energies  of the scattered electron $E'_l$ and the recoil proton $E'_p$ can be expressed
also through $E_l$ and $Q^2$:
\be
E'_p =\frac{Q^2}{2M} + M~, \;\;E'_l = E_l  - \frac{Q^2}{2M}~. \label{E'p E'l}
\ee
Experimentally observed cross section  $d\sigma_ {exp}$ usually is presented as
\be
 {d\sigma_ {exp}} =  \, {d\sigma_B} (1+ \delta)~,
 \label{dsigma exp}
\ee
where the radiative correction $ \delta$ is given by the sum of the correction $\delta_{virt}$ accounting contributions of higher orders in the electromagnetic coupling $\alpha$ to  the  elastic scattering cross section and  the correction  $\delta_{real}$ due to inelastic processes admitted by  experimental conditions:  $ \delta = \delta_{virt} + \delta_{real} $.  In the one-loop approximation the virtual corrections $ \delta_ {virt} $ is determined by  the interference of the Born amplitude with the one-loop amplitudes, and the real correction $ \delta_ {real} $ is  related to the  one-photon emission.

Taking separately, the virtual and real contributions has no sense, neither from theoretical, nor from experimental point of view; in the first case -- because of infrared divergencies, in the second -- because only processes with emission of any number of sufficiently soft photons are experimentally observed.

The most important virtual correction is so called vertex correction. In the one-loop approximation it is well known (see, for example, \cite{BLP})
\be
\delta_{vertex}^{e} = \frac{\alpha}{\pi} \left[-\left(\ln\left(\frac{Q^2}{m^2}\right)-1\right)\, \ln\left(\frac{m^2}{\lambda^2}\right) - \frac{1}{2}\,\ln^2\left(\frac{Q^2}{m^2}\right) + \frac{3}{2} \ln\left(\frac{Q^2}{m^2}\right) +\frac{\pi^2}{6}- 2 \right]~. \label{vertex}
\ee
Here $\lambda$ is the "photon mass" introduced for regularization of the infrared divergencies.

Photons emitted  in the scattering process by an electron can be conveniently divided, as is usually done, into soft, non-affecting the elastic kinematics of the process, and hard. The soft photons can be defined as those which  have energy less than  $\omega_0$ (with a sufficiently small  $\omega_0$) in some reference frame. For them, the factorization theorem \cite{Yennie:1961ad} can be used and their contribution  $\delta_{soft}^{e}$ to  $\delta_{real}$ can be found in any frame  using the integrals calculated in \cite{'tHooft:1978xw}. If the limitation  on the  photon energy $\omega < \omega_0$ is imposed in the rest proton system, then  (see, for example, \cite{Gerasimov:2015aoa})
\[
\delta_{soft}^{e} =
\frac{\alpha}{\pi}
\left[
  \left(\ln\left(\frac{Q^2}{m^2}\right) - 1\right)\, \ln\left(\frac{4\omega_0^2}{\lambda^2}\right)
  - \frac{1}{4} \, \ln^2\left(\frac{4E_l^2}{m^2}\right)
  - \frac{1}{4} \, \ln^2\left(\frac{4E_l'{}^2}{m^2}\right)
\right.
\]
\be
\left.
  + \frac{1}{2} \ln^2\left(\frac{4 E_l E_l'}{Q^2}\right)
  + \ln\left(\frac{4E_lE_l'}{m^2}\right)
  + \text{Li}_2\left(1 - \frac{Q^2}{4E_l E_l'}\right)
  - \frac{\pi^2}{3}
\right]
\ee
The sum of the vertex and the  soft photon  corrections is free from infrared singularities
\[
    \delta_{vertex + soft}^{e} = \frac{\alpha}{\pi}
    \left[
      -\left(\ln\left(\frac{Q^2}{m^2}\right)-1\right)\, \ln\left(\frac{E_l E_l'}{\omega_0^2}\right)
      +\frac32 \,\ln\left(\frac{Q^2}{m^2}\right) - 2
    \right.
\]
\be
    \left.
      - \frac{1}{2}\ln^2\left(\frac{E_l}{E_l'}\right)
      + \text{Li}_2\left(1 - \frac{Q^2}{4E_l E_l'}\right)
      - \frac{\pi^2}{6}
    \right]~,  \label{soft + hard}
\ee
so that the hard photon contribution to $\delta_{real}$ can be calculated with  zero photon mass. The cancellation of the infrared singularities in (\ref{soft + hard})  is well known. It is a consequence of the general statement \cite{Yennie:1961ad} about the cancellation of infrared singularities in the sum of virtual corrections and corrections due to the emission of soft photons.  However, the sum (\ref{soft + hard})  contains $\ln\left(\frac{Q^2}{m^2}\right)$, i.e. so called collinear singularities. In general, the terms with $\ln\left(\frac{Q^2}{m^2}\right)$ remain also after account of contributions of hard photons, because the theorem on the cancellation of  collinear singularities \cite{Kinoshita:1962ur}-\cite{Lee:1973} evidently can not be applied to radiative corrections to electron-proton scattering.
However, in the experimental conditions of Ref.~\cite{Vorobev} these terms disappear.
It is amusing enough  that their cancellation  may be related to the Kinoshita--Lee--Nauenberg theorem \cite{Kinoshita:1962ur}-\cite{Lee:1973}. It will be explained  in the next section. And here we will  demonstrate  it using less advanced arguments.

We have to add to (\ref{soft + hard}) the correction from  the  hard photon emission. With logarithmic accuracy, this correction has a clear physical interpretation. It consists of two parts, corresponding to photon emission by initial and final electrons. Both of them can be calculated using the quasi-real electron method \cite{Baier:1973ms}.
For the final electron emission it gives
\be
\frac{\omega \; d\sigma^{f.e.e.}}{d^3k}  = \frac{\alpha}{4\pi^2} \left( \frac{E'^2_l+(E'_l -\omega)^2}{\omega\;E'_l (kl')} -\frac{m^2}{(kl')^2}\frac{(E'_l -\omega)}{E'_l}\right)d\sigma_B~,  \label{f.e.-1}
\ee
where $k$ and $\omega$ are  the  photon 4-momentum and energy, $l'$ and $E'_l$ are the  4-momentum and energy of the final electron in the elastic process,  and   $d\sigma_B$ is the cross section of the elastic process.
With the logarithmic accuracy,  the upper limit of the integration over photon emission angle should be taken equal to the scattering angle in the elastic process, so that we obtain with this accuracy
\be
\frac{x\;d\sigma^{f.e.e.}}{dx}  =\frac{\alpha}{2\pi}\ln\frac{Q^2}{m^2}
\left(1+(1-x)^2\right) d\sigma_B~, \label{f.e.-2}
\ee
where $x=\omega/E'_l$. Integration over $x$ can  be performed from $\omega_0/E'_l$ to 1,  that gives
\be
d\sigma^{f.e.e.} = \frac{\alpha}{\pi}\ln\frac{Q^2}{m^2}\left(\ln\frac{E'_l}{\omega_0}-\frac34\right) d\sigma_B~.  \label{f.e.-3}
\ee

Comparing it with (\ref{soft + hard})
one sees that the contribution (\ref{f.e.-3}) cancels  in (\ref{soft + hard}) the collinear singular  terms with the coefficient $\ln\frac{E'_l}{\omega_0}$ and   a half of the terms  with the  energy independent coefficient.

For the correction  due to radiation of the initial  electron we have, with the same  accuracy
\be
\frac{\omega \; d\sigma^{i.e.e.}}{d^3k}  = \frac{\alpha}{4\pi^2} \left( \frac{E^2_l+(E_l -\omega)^2}{\omega\;E_l (kl)} -\frac{m^2}{(kl)^2}\frac{(E_l -\omega)}{E_l}\right)d\sigma_B\Big|_{\vec{l} \rightarrow \vec{l}-\vec{k}}~,    \label{i.e.-1}
\ee
and after integration with the logarithmic accuracy over emission angles
\[
d\sigma^{i.e.e.}  = \frac{\alpha}{2\pi}\ln\frac{Q^2}{m^2}\int_{\omega_0/E_l}^1\frac{dx}{x}
\left(1+(1-x)^2\right)d\sigma_B\Big|_{\vec{l} \rightarrow \vec{l}(1-x)}~,\label{i.e.-2}
\]
with $x=\omega/E_l$.

Emission of a photon by an initial electron changes its energy, and thus cross sections of processes  initiated by the electron after radiation (quasi-real electron). This is the reason why  virtual corrections are not cancelled by the real ones in the  most experiments. But in the proposed experiment \cite{Vorobev}, where the differential in the momentum transferred to the final proton, $d\sigma/(dQ^2) $,    is supposed to be measured  in the region
\be 
 Q^2 \ll E_l^2~, \;\; Q^2 \ll  E_l M~, \label{region}
\ee 
one has $E'_l \simeq E_l$ 
and
\be
\frac{d\sigma_B}{dQ^2} \simeq  \frac{4\pi\alpha^2}{Q^4}f_1^2(Q^ 2),
 \label{born-1}
\ee
i.e. $d\sigma_B/(dQ^2)$ does not depend on  $E_l $  at fixed $Q^2$, so that $d\sigma^{i.e.e.}$ equal to  $d\sigma^{f.e.e.}$   and in the sum  they cancel the terms with $\ln\frac{Q^2}{m^2}$ in (\ref{soft + hard}).

In fact,  the cancellation   of the virtual and real corrections   turns out to be even stronger. It occurs that in the one-loop approximation the cancellation   takes place not only with logarithmic accuracy, and  the terms not having the collinear singularities  (constant terms) are cancelled as well. Of course, such cancellation can not be proved by some approximate method and requires more strict approach. It can be seen using exact  results of Ref. \cite{Baier:197} for a photon emission in collisions of muons with electrons.  Making the necessary replacements and substitutions one can obtain  for the cross section of photon emission by electron  in the  experimental conditions discussed above (detailed output will be given elsewhere \cite{FG:2019})
\[
d\sigma^{e.e.}= \frac{4 \alpha^3}{Q^4}f_1^2(Q^ 2)\int_{\omega_0/E_l}^{1}\frac{dx}{x} \Biggl[ \bigl(1+(1-x)^2\bigr)\ln\frac{Q^2}{m^2} -2(1-x)\Biggr]
\]
\be
=\frac{d\sigma_B}{dQ^2}\frac{\alpha}{\pi}  \Biggl[ 2\left(\ln\frac{Q^2}{m^2}-1\right)\ln\frac{E_l}{\omega_0} -\frac32 \,\ln\frac{Q^2}{m^2} + 2 \Biggr]~. \label{e.e.-1}
\ee
Comparing it with (\ref{soft + hard}) at $E'_l \simeq E_l, \;\; Q^2 \ll E_l^2$,
one sees that the contribution (\ref{e.e.-1})  cancels it completely.

\section{Higher order corrections}

The  cancellation demonstrated above   is not restricted  by the one-loop approximation, and  takes place also in higher orders, at least with  logarithmic accuracy. It can be  shown using the parton picture,  developed for the theoretical description of the deep inelastic electron-proton scattering \cite{Gribov:1972ri}-\cite{Altarelli:1977zs}, but applying it to  "deep inelastic proton-electron scattering".    

The cross section of electron-proton scattering with radiative corrections  due only to electron interaction  can be considered as inclusive proton-electron scattering cross section.  At fixed $Q^2$ it   can be written  with  logarithmic accuracy   in terms of the "parton distribution functions" $f_e^e(x, Q^2)$ and  $f_e^{\bar{e}}(x, Q^2)$.
More precisely, the cross section of the deep inelastic proton-electron scattering 
can be  written as
\be
(2\pi)^32E'_p\frac{d^3\sigma}{d^3p'}
=\frac{\pi e^4}{Q^4}\frac{1}{\sqrt{(pl)^2-m^2M^2}}J^{\mu\nu}(p, p')
W_{\mu\nu}(l, q)~, \label{cross section = L munu W munu}
\ee
where $J^{\mu\nu}(p, p')$ is the proton current tensor
\be
J^{\mu\nu}(p, p') = \overline{\sum}_{pol}J^\mu J^{*\nu}~,
\ee
 $\overline{\sum}_{pol}$  means summation over final polarizations and averaging over initial ones,
 \be
 J^\mu = u(p')\,\left(f_1(Q^2)\,\gamma^\mu + f_2(Q^2)\,\frac{[\gamma^\mu, \gamma^\nu]q_\nu}{4M}\right)\,u(p)~,
 \ee
 and $W_{\mu\nu}(l, q)$ is the deep inelastic scattering tensor,
 \be
 {W}_{\mu\nu}(l, q)=\frac{1}{4\pi} \overline{\sum}_X\langle
 l|j_\nu^{(e)}(0)|X\rangle~\langle
 X|j_\mu^{(e)}(0)|l\rangle~(2\pi)^4\delta(q+l-p_X)~.
\ee
Here $|l>$ is the initial electron state, $X$ is any state which can be produced in photon-electron collisions, $\overline{\sum}_X$ means averaging over initial electron polarizations and   summation over discrete and integration over continuous variables of $X$,
$j_\mu^{(e)}(x)$ is the electron electromagnetic current  current operator.
 Taking into account the conservation of the current, one can represent $W_{\mu\nu}$ in the form
\be
{W}^{\mu\nu}(l, q)=-\left(g^{\mu\nu}-\frac{q^\mu q^\nu}{q^2}\right)F_1(x, Q^2)+\frac{1}{(l q)}\left(l^{\mu}-\frac{(l q)}{q^2}q^\mu\right)\left(l^{\nu}-\frac{(lq)}{q^2 }q^\nu\right)F_2(x, Q^2)~,
\ee
where $Q^2=-q^2~, \;\; x=Q^2/(2(l q))$.
Calculating the tensor $J^{\mu\nu}$
\be
J^{\mu\nu} =G_M^2(Q^2)\left(g_{\mu\nu}q^2-{q_\mu q_\nu}\right)+\frac{Q^2G_M^2(Q^2) +4M^2 G_E^2(Q^2)}{4M^2+Q^2}P^\mu P^\nu~,
\ee
where $P =p+p'$,  $G_E(Q^2)$ and $G_M(Q^2)$ are proton electric and magnetic form factors, performing tensor convolution and using
\be
\frac{d^3p'}{2E'_p} =\frac{\pi}{4}\frac{Q^2 dQ^2 dx}{x^2\sqrt{(pl)^2-m^2M^2}}~,
\ee
we obtain
\[
\frac{d\sigma}{dQ^2 dx} = \frac{\pi\alpha^2}{2x^2Q^2((pl)^2-m^2M^2)}
\left[(2Q^2G_M^2 -4M^2G_E^2)F_1 +\right.
\]
\be\left.
 \left(-G_M^2(m^2 Q^2 +(lq)^2)+\frac{Q^2G_M^2 +4M^2G_E^2}{4M^2+Q^2}(Pl)^2  \right)\frac{F_2}{(lq)} \right]~.
\ee
where  
\be
(lq) =\frac{Q^2}{2x}~,\;\;   (Pl) = 2M E_l - \frac{Q^2}{2x}~. 
\ee
The region of variation of $x$ at fixed $Q^2$ is determined by conditions $M_X^2 \ge m^2$ and $(lq) \le E_lq_0 +\sqrt{E_l^2-m^2}\sqrt{q_0^2+Q^2}$   with $q_0=M-E'_p = -Q^2/(2M)$, i.e 
\be
1\ge x\ge \frac{MQ^2}{\sqrt{E_l^2-m^2} \sqrt{Q^2(4M^2 +Q^2)} -E_l Q^2}~. 
\ee  
Till now we did not make any approximation. Using  the conditions (\ref{region}) of the proposed experiment  and the Callan-Gross  relation \cite{Callan:1969uq} $F_2 =2x F_1$, 
we obtain  
\be
\frac{d\sigma}{dQ^2} = \frac{4\pi\alpha^2}{Q^4}f^2_1(Q^2)
\int_{x_0}^1\frac{dx}{x} F_2(x, Q^2)~, \label{sigma - F2}
\ee
where 
\be
x_0 =\frac{Q}{2E_l}\ll 1~. 
\ee
In the parton picture \cite{Gribov:1972ri}-\cite{Altarelli:1977zs}, which can be used with the logarithmic accuracy, the structure functions are expressed through the parton distributions in the initial  electron.   
The charged partons in this case are electrons and positrons, so that
\be 
F_2 = x (f_e^e  + f_e^{\bar e})~, \label{F2=f+f}
\ee
where $f_e^e$ and $f_e^{\bar e}$ are  the electron and positron distributions in the initial electron.  

The  positron distribution $f_e^{\bar e}$ appears  in the two-loop approximation, so that  in the one-loop approximation only $f_e^e $ does contribute. Moreover, in this approximation $f_e^e $  is equal to the valent electron distribution,  $f_e^v $,  which is  not singular at small $x$, so that the lower limit $x_0$ of the  integration in (\ref{sigma - F2}) can be taken equal to zero. Therefore  the cancellation of the logarithmic radiative corrections discussed in the previous section has a simple  explanation  in the language of parton distributions: it takes place because
\be
\int_0^1 dx f_e^v (x, Q^2) =1~ \label{charge conservation}
\ee
regardless of the value of $Q^2$. Remind that (\ref{charge conservation}) is the consequence of the charge conservation. 

Surprisingly  enough, this cancellation can be explained with the help of
the Kinoshita--Lee--Nauenberg theorem.  Of course, this theorem can not be applied directly to the electron-proton scattering process. But it can be applied to completely inclusive production of electrons, positrons and photons by the photon with  the  virtuality $Q^2$.   Using the famous  "reciprocal relation"  \cite{Gribov:1972ri}-\cite{Altarelli:1977zs}, \cite{Bukhvostov:1974uu, Fishbane:1973ji},   which means in our case equality  of the parton distribution function  $f_e^e(x, Q^2)$  and   the "parton fragmentation function" $\bar{f}_e^e(x, Q^2)$ for the inclusive cross section of electron production  by the photon with  the  virtuality $Q^2$,  one  can connect two processes and to prove  the cancellation (for details see \cite{FG:2019}).  

Starting from the two-loop approximation the situation becomes not so simple. 
The radiative corrections depend strongly on what is really measured, and if  the  fully inclusive cross  section  (\ref{sigma - F2}) is measured, they become large. 
The reason is that in this approximation the target proton can interact not only with the scattering  electron, but also  with  one of  components of the electron-positron pair produced  by this electron.  In this case the total cross section for interaction  of the virtual photon emitted by the target proton with incoming electron    does not decrease  with energy, in contrast with the one-loop approximation, and the large contribution come from the region of small $(lq)$, or small $x$ in (\ref{sigma - F2}). On the language of the parton distributions it means that  both $f_e^{e}(x, Q^2)$ and $f_e^{\bar e}(x, Q^2)$ becomes singular at $x=0$ and  the lower limit $x_0$ of the  integration in (\ref{sigma - F2}) can not be taken equal to 0.  Evidently, such experimental conditions are not the best one. It seems that more preferable are the conditions at which production  of electron-positron pairs is forbidden. In this case in Eq. (\ref{F2=f+f}) only $f_e^v$ does contribute, and due to its property (\ref{charge conservation})  the major pieces of the  radiative corrections  cancel also in higher orders of perturbation theory (for details see \cite{FG:2019}).

\section{Conclusion}
It turns out that the setting of the experiment suggested by A.A. Vorobev \cite{Vorobev} has an interesting feature -- cancellation of main radiative corrections. We showed it in this paper by several methods and with various accuracy. The most simple and physically transparent is the quasi-real electron method \cite{Baier:1973ms} which can be used in the one-loop approximation and has logarithmic accuracy.  It turns out, however,  that in the one-loop approximation the cancellation  of the virtual and real radiative corrections is not restricted by the logarithmic accuracy, but takes place  for the terms not having  the collinear divergence (constant terms) as well. It was shown with the help of  the exact results of Ref. \cite{Baier:197} for a photon emission in collisions of muons with electrons. Moreover, it turns out, that for the  experimental conditions  at which production  of electron-positron pairs is forbidden, the cancellation   is not restricted by the  one-loop approximation and with the logarithmic accuracy  it takes place in the higher orders of perturbation theory  also. It was demonstrated using the parton distribution  method \cite{Gribov:1972ri}-\cite{Altarelli:1977zs} developed for the theoretical description of  the deep inelastic electron-proton scattering. It is interesting that the famous reciprocal relation \cite{Gribov:1972ri}-\cite{Altarelli:1977zs}, \cite{Bukhvostov:1974uu, Fishbane:1973ji} permits to prove the cancellation of the terms with $\ln\frac{Q^2}{m^2}$ in the one-loop radiative corrections with the help of the Kinoshita--Lee--Nauenberg theorem \cite{Kinoshita:1962ur}-\cite{Lee:1973}.

\end{document}